\documentclass{moriond}

\usepackage{amsmath}

\usepackage{tikzscale}
\usepackage{tikz-feynman}

\usepackage{ifthen} 
\newboolean{uprightparticles}
\setboolean{uprightparticles}{false} 

\usepackage{lineno}
\usepackage{fancyhdr}
\def\Journal#1#2#3#4{{#1} {\bf #2}, #3 (#4)}

\def\NIM{\em Nucl. Instrum. Methods}

\def\PRL{\em Phys. Rev. Lett.}
\def\PRD{{\em Phys. Rev.} D}

\def\be{\begin{equation}}
\def\ee{\end{equation}}
\def\bea{\begin{eqnarray}}
\def\eea{\end{eqnarray}}

\usepackage{xspace} 
\usepackage{upgreek}

\def\MagUp {\mbox{\em Mag\kern -0.05em Up}\xspace}

\ifthenelse{\boolean{uprightparticles}}%
{

 \def\Pmu         {\ensuremath{\upmu}\xspace}                 
 \def\Pnu         {\ensuremath{\upnu}\xspace}                 
                  
 \def\Ppi         {\ensuremath{\uppi}\xspace}

 \def\Ptau        {\ensuremath{\uptau}\xspace}

 \def\PDelta      {\ensuremath{\Delta}\xspace}                 
 \def\PXi         {\ensuremath{\Xi}\xspace}                 
 \def\PLambda     {\ensuremath{\Lambda}\xspace}                 
 \def\PSigma      {\ensuremath{\Sigma}\xspace}                 
 \def\POmega      {\ensuremath{\Omega}\xspace}                 
 \def\PUpsilon    {\ensuremath{\Upsilon}\xspace}
 \let\oldPi\Pi
 \def\PPi         {\ensuremath{\oldPi}\xspace}

 \def\PB      {\ensuremath{\mathrm{B}}\xspace}                 
                  
 \def\PD      {\ensuremath{\mathrm{D}}\xspace}

 \def\PK      {\ensuremath{\mathrm{K}}\xspace}

 \def\Pi      {\ensuremath{\mathrm{i}}\xspace}

 \def\Ps      {\ensuremath{\mathrm{s}}\xspace}

 \def\thebaroffset{0.0em}
}
{

 \def\Pmu         {\ensuremath{\mu}\xspace}                 
 \def\Pnu         {\ensuremath{\nu}\xspace}                 
                  
 \def\Ppi         {\ensuremath{\pi}\xspace}

 \def\Ptau        {\ensuremath{\tau}\xspace}

 \mathchardef\PDelta="7101
 \mathchardef\PXi="7104
 \mathchardef\PLambda="7103
 \mathchardef\PSigma="7106
 \mathchardef\POmega="710A
 \mathchardef\PUpsilon="7107
 \mathchardef\PPi="7105
                  
 \def\PB      {\ensuremath{B}\xspace}                 
                  
 \def\PD      {\ensuremath{D}\xspace}

 \def\PK      {\ensuremath{K}\xspace}

 \def\Pi      {\ensuremath{i}\xspace}

 \def\Ps      {\ensuremath{s}\xspace}

 \def\thebaroffset{0.18em}
}
\newcommand{\offsetoverline}[2][\thebaroffset]{\kern #1\overline{\kern -#1 #2}}%

\makeatletter
\ifcase \@ptsize \relax
  \newcommand{\miniscule}{\@setfontsize\miniscule{4}{5}}
\or
  \newcommand{\miniscule}{\@setfontsize\miniscule{5}{6}}
\or
  \newcommand{\miniscule}{\@setfontsize\miniscule{5}{6}}
\fi
\makeatother

\DeclareRobustCommand{\optbar}[1]{\shortstack{{\miniscule (\rule[.5ex]{1.25em}{.18mm})}
  \\ [-.7ex] $#1$}}

\def\mun        {{\ensuremath{\Pmu^-}}\xspace} 

\def\taup       {{\ensuremath{\Ptau^+}}\xspace}
\def\taum       {{\ensuremath{\Ptau^-}}\xspace}

\def\neu        {{\ensuremath{\Pnu}}\xspace}
\def\neub       {{\ensuremath{\overline{\Pnu}}}\xspace}

\def\neumb      {{\ensuremath{\neub_\mu}}\xspace}
\def\neut       {{\ensuremath{\neu_\tau}}\xspace}
\def\neutb      {{\ensuremath{\neub_\tau}}\xspace}

\def\squark    {{\ensuremath{\Ps}}\xspace}

\def\pion   {{\ensuremath{\Ppi}}\xspace}
\def\piz    {{\ensuremath{\pion^0}}\xspace}
\def\pip    {{\ensuremath{\pion^+}}\xspace}
\def\pim    {{\ensuremath{\pion^-}}\xspace}

\def\kaon    {{\ensuremath{\PK}}\xspace}

\def\KorKbar {\kern \thebaroffset\optbar{\kern -\thebaroffset \PK}{}\xspace}

\def\Km      {{\ensuremath{\kaon^-}}\xspace}

\def\D       {{\ensuremath{\PD}}\xspace}

\def\DorDbar {\kern \thebaroffset\optbar{\kern -\thebaroffset \PD}\xspace}
\def\Dz      {{\ensuremath{\D^0}}\xspace}

\def\Dp      {{\ensuremath{\D^+}}\xspace}
\def\Dm      {{\ensuremath{\D^-}}\xspace}

\def\DpDm    {\ensuremath{\Dp {\kern -0.16em \Dm}}\xspace}
\def\Dstar   {{\ensuremath{\D^*}}\xspace}

\def\Dstarp  {{\ensuremath{\D^{*+}}}\xspace}
\def\Dstarm  {{\ensuremath{\D^{*-}}}\xspace}

\def\B       {{\ensuremath{\PB}}\xspace}
\def\Bbar    {{\ensuremath{\offsetoverline{\PB}}}\xspace}

\def\BorBbar {\kern \thebaroffset\optbar{\kern -\thebaroffset \PB}\xspace}
\def\Bz      {{\ensuremath{\B^0}}\xspace}
\def\Bzb     {{\ensuremath{\Bbar{}^0}}\xspace}
\def\Bd      {{\ensuremath{\B^0}}\xspace}

\def\BdorBdbar {\kern \thebaroffset\optbar{\kern -\thebaroffset \Bd}\xspace}

\def\Bs      {{\ensuremath{\B^0_\squark}}\xspace}

\def\BsorBsbar {\kern \thebaroffset\optbar{\kern -\thebaroffset \Bs}\xspace}

\def\Y#1S{\ensuremath{\PUpsilon{(#1S)}}\xspace}

\def\LorLbar     {\kern \thebaroffset\optbar{\kern -\thebaroffset \PLambda}\xspace}

\def\to                 {\ensuremath{\rightarrow}\xspace}

\def\AT#1     {\ensuremath{A_{\mathrm{T}}^{#1}}\xspace}           

\def\C#1      {\ensuremath{\mathcal{C}_{#1}}\xspace}                       
\def\Cp#1     {\ensuremath{\mathcal{C}_{#1}^{'}}\xspace}                    
\def\Ceff#1   {\ensuremath{\mathcal{C}_{#1}^{\mathrm{(eff)}}}\xspace}        
\def\Cpeff#1  {\ensuremath{\mathcal{C}_{#1}^{'\mathrm{(eff)}}}\xspace}       
\def\Ope#1    {\ensuremath{\mathcal{O}_{#1}}\xspace}                       
\def\Opep#1   {\ensuremath{\mathcal{O}_{#1}^{'}}\xspace}                    

\newcommand{\aunit}[1]{\ensuremath{\text{\,#1}}}       

\newcommand{\tev}{\aunit{Te\kern -0.1em V}\xspace}
\newcommand{\gev}{\aunit{Ge\kern -0.1em V}\xspace}
\newcommand{\mev}{\aunit{Me\kern -0.1em V}\xspace}
\newcommand{\kev}{\aunit{ke\kern -0.1em V}\xspace}
\newcommand{\ev}{\aunit{e\kern -0.1em V}\xspace}
 
\newcommand{\mevc}{\ensuremath{\aunit{Me\kern -0.1em V\!/}c}\xspace}
\newcommand{\gevc}{\ensuremath{\aunit{Ge\kern -0.1em V\!/}c}\xspace}
\newcommand{\mevcc}{\ensuremath{\aunit{Me\kern -0.1em V\!/}c^2}\xspace}
\newcommand{\gevcc}{\ensuremath{\aunit{Ge\kern -0.1em V\!/}c^2}\xspace}

\def\fb   {\ensuremath{\aunit{fb}}\xspace}
\def\invfb   {\ensuremath{\fb^{-1}}\xspace}

\def\gsim{{~\raise.15em\hbox{$>$}\kern-.85em
          \lower.35em\hbox{$\sim$}~}\xspace}
\def\lsim{{~\raise.15em\hbox{$<$}\kern-.85em
          \lower.35em\hbox{$\sim$}~}\xspace}

\def\sPlot{\mbox{\em sPlot}\xspace}

\def\tell1  {TELL1\xspace}
\def\ukl1   {UKL1\xspace}

\newcommand{\eg}{\mbox{\itshape e.g.}\xspace}

\newcommand{\lhcborcid}[1]{\href{https://orcid.org/#1}{\hspace*{0.1em}\raisebox{-0.45ex}{\includegraphics[width=1em]{figs/orcidIcon.pdf}}}}

\newcommand{\Photo}{}

\begin{document}

\vspace*{4cm}
\title{$b\to c l \overline{\nu}$ decays at LHCb}

\author{C. Chen \\ (on behalf of the LHCb Collaboration)}

\address{Aix Marseille Univ, CNRS/IN2P3, CPPM,\\
  163, avenue de Luminy - Case 902, 13288 Marseille, France}

\maketitle\abstracts{
This report introduces two recent measurements of semileptonic $b$-hadron decays at the LHCb experiment, including a test of Lepton Flavour Universality~(LFU) using $\Bzb\to D^{(*)+} l^- \neub_l$ decays where $l\in\{\mu,\tau\}$, and a study of the \Dstarp longitudinal polarisation in the $\Bzb\to\Dstarp\taum\neutb$ decay. 
With the inclusion of the new results of the LFU ratios, the world average on $R(D)$ and $R(\Dstar)$, still shows a tension over three standard deviations from the SM prediction, while the measured \Dstarp longitudinal polarisation is compatible with the SM value.
}

\section{Introduction}

Semileptonic decays of $b$-hadrons in the Standard Model~(SM) proceed  via the $b\to cl\neub$ transition mediated by the $W$ boson at the tree level.
Such decays present valuable opportunities to search for New Physics~(NP) beyond the SM. 
Additional particles not included in the SM
, \eg charged Higgs and leptoquarks~\cite{Crivellin:2015hha}, 
might also contribute to the $b\to c l\neub$ process alongside the $W$ boson,
leading to deviations of physics observables from SM expectations. 
Precise measurements of these observables are crucial for testing the SM and exploring potential NP. 
Notable examples of these observables are Lepton Flavour Universality~(LFU) ratios, defined as the branching-fraction ratio of two semileptonic decays involving leptons in different generations, and angular coefficients that characterise the differential decay width across the phase space.


There is a long-standing tension in the world averages of the LFU ratios, $R(D^{(*)})\equiv \mathcal{B}(\Bbar\to D^{(*)}\taum\neutb)/\mathcal{B}(\Bbar\to D^{(*)}\mu^-\neub_\mu)$,
which deviate from the SM predictions by over three standard deviations~($\sigma$)~\cite{HFLAV21}. This discrepancy has generated considerable interest within the particle physics community to explore whether this tension is due to the NP. 
This report presents two recent LHCb measurements that contribute to this effort: the simultaneous measurement of $R(D^{(*)+})$~\cite{LHCb-PAPER-2024-007} and the analysis of the \Dstarp longitudinal polarisation in the $\Bzb\to\Dstarp\taum\neutb$ decay~\cite{LHCb-PAPER-2023-020}.
These analyses benefit from the dedicated design and the excellent performance of the LHCb detector for reconstructions of $b$-hadron decays. 

\section{Simultaneous measurement of $R(\Dstarp)$ and $R(\Dp)$}

Using proton-proton~($pp$) collision data corresponding to an integrated luminosity of $2\invfb$ collected by the LHCb experiment during the periods 2015-2016, the LFU ratios $R(\Dstarp)$ and $R(\Dp)$ are measured as follows~\cite{LHCb-PAPER-2024-007}:
\begin{equation}
    R(D^{(*)+}) \equiv {  {{\cal B}(\Bzb\to \D^{(*)+}\taum \neutb)}  \over  {{\cal B}(\Bzb\to \D^{(*)+}\mun \neumb)} }\, .
\end{equation}
The tauoinc and muonic decays are denoted as signal and normalisation channels, respectively.
This is the first LHCb analysis measuring $R(D^{(*)})$ with the \Dp meson.
The \Dstarp meson decays into the \Dp meson and an unreconstructed neutral pion or photon, followed by the $\Dp\to\Km\pip\pip$ decay. The \taum lepton decays 
 into $\mun\neumb\neut$ where neutrinos escape from detection. Consequently, the only reconstructed final state in both signal and normalisation channels is $\Dp\mun$, enabling direct and simultaneous determination of $R(D^{(*)+})$ with the same data sample.

Simulation samples are used to optimise selections, construct template probability density functions~(PDF) and calculate efficiencies. This analysis marks the first use of the Tracker-Only simulation at LHCb, where only the detector responses from the tracking system are simulated. The effects of the particle identification~(PID), calorimeter and muon systems are emulated during the offline analysis. This Tracker-Only configuration speeds up the simulation process by a factor of eight compared to the full simulation, allowing for the production of large amounts of simulated events. This capability is crucial for reducing relevant systematic uncertainties which are typically one of the main constraints on the precision of $R(D^{(*)})$ results at LHCb. 

The selection of signal candidates starts with triggers that persist the muon kinematic features to distinguish tauonic from muonic decays, followed by kinematic, track-quality and PID requirements for final-state particles.
The \Dp decay vertex must be significantly displaced from the $pp$ collision vertex to reduce background directly from $pp$ collisions. 
Several Boosted Decision Trees~(BDT)~\cite{BDT} are employed to suppress backgrounds from fake tracks, accidental track combinations forming fake \Dp candidates, and partially reconstructed decays with missing neutral or charged objects in addition to $\Dp\mun$. The residual fake \Dp contribution is statistically subtracted using the \sPlot technique~\cite{sPlot} with $m(\Km\pip\pip)$ as the discriminating variable.

Signal and normalisation yields are determined via a three-dimensional binned-likelihood fit to the squared four-momentum transferred to the lepton system~($q^2$), lepton energy in the \Bzb rest frame~($E_l^*$) and the invariant-mass of unreconstructed particles~($m_{\rm miss}^2$). 
The PDFs of the components are parameterised as histogram templates, which are derived from simulation for $B$ semileptonic and double charm decays, and from data for backgrounds due to $\mu$ misidentification and random $\Dp\mun$ combinations.
The form factors for the signal and normalisation decays are parameterised using the BGL formalism~\cite{BGL}, while for semileptonic $B$ decays involving higher excited $D$ states, the BLR formalism is used~\cite{BLR}. The form-factor parameters are allowed to vary in the fit.
The fit is performed simultaneously across four data samples, each enriched with specific decays. The signal sample, comprising $\Dp\mun$ candidates, primarily contains signal and normalisation decays. Three control samples, which additionally include either one or two charged pions, or a charged kaon, predominantly consist of feature contributions from feed down of semileptonic $B$ decays involving higher excited $D$ states and double charm decays with two charm mesons.
The three control samples provide crucial constraints on the background fractions in the signal sample. 

\begin{figure}[tbp]
    \centering
\begin{minipage}[t]{.42\textwidth}
    \centering
    \includegraphics[width=1.0\linewidth]{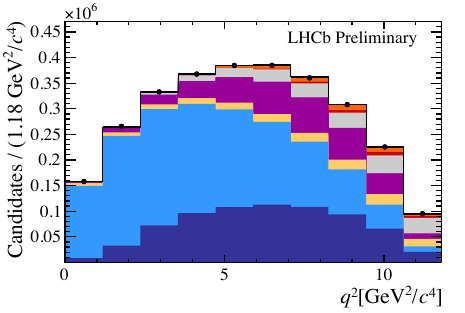}
\end{minipage}
\begin{minipage}[t]{.42\textwidth}
    \centering
    \includegraphics[width=1.0\linewidth]{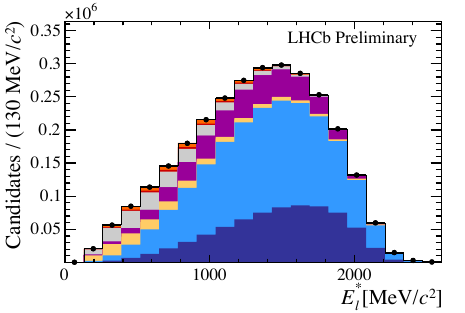}
\end{minipage}
\begin{minipage}[t]{.42\textwidth}
    \centering
    \includegraphics[width=1.0\linewidth]{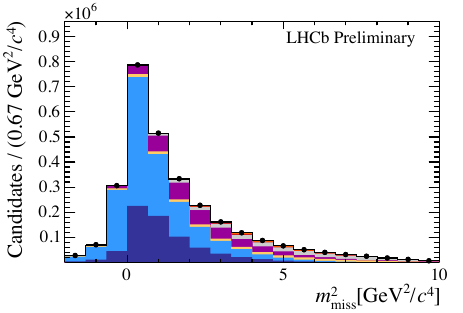}
\end{minipage}
\begin{minipage}[t]{.42\textwidth}
    \vspace{-0.63\linewidth}
    \hspace{0.25\linewidth}
    \includegraphics[width=0.5\linewidth]{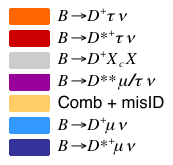}
\end{minipage}
    \caption{Distributions of the three fit variables in the signal sample with the fit result overlaid.}
    \label{fig:rdst_fit}
\end{figure}

Distributions of the three fit variables in the signal sample are shown in Fig.~\ref{fig:rdst_fit} with the fit results overlaid. There is a good agreement observed between the fit results and data distributions. With the efficiencies of the signal and normalisation channels determined from simulation, the LFU ratios are measured to be 
\begin{equation}
R\left(D^{+}\right) =0.249 \pm 0.043 \pm 0.047,  
\quad
R\left(D^{*+}\right) =0.402 \pm 0.081 \pm 0.085, 
\label{eq:rdst_result}
\end{equation}
with a correlation factor of $\rho=-0.39$. Here, the first uncertainties are statistical and the second systematic. The primary sources of systematic uncertainties are the form-factor parameterisation and background modelling. The limited size of simulation samples is also a main source contributing to the uncertainty on $R\left(D^{*+}\right)$.
The $R\left(D^{+}\right)$ and $R\left(D^{*+}\right)$ results shown in Eq.~\ref{eq:rdst_result} are compatible with the world average of all relevant measurements and the SM predictions~\cite{HFLAV21}. With these results included, the updated world average still exhibits a tension with the SM expectation at a $3\sigma$ level.

\section{Measurement of the \Dstarp longitudinal polarisation}

In the $\Bzb\to\Dstarp\taum\neutb$ decay, the \Dstarp longitudinal polarisation factor $F_L^{D^*}$ is defined as 
\begin{equation}
F_L^{D^*} = \frac{a_{\theta_D}(q^2)+c_{\theta_D}(q^2)}{3 a_{\theta_D}(q^2)+c_{\theta_D}(q^2)}.
\end{equation}
The parameters $a_{\theta_D}(q^2)$ and $c_{\theta_D}(q^2)$ are the coefficients characterising the sizes of unpolarised and polarised components in the differential decay width 
\begin{equation}
\frac{\mathrm{d}^2 \Gamma}{\mathrm{d} q^2 \mathrm{d} \cos \theta_D} = a_{\theta_D}(q^2)+c_{\theta_D}(q^2) \cos ^2 \theta_D. 
\end{equation}
In the case that $\Dstarp$ decays into $\Dz\pip$, $\theta_D$ is the angle between the \Dz momentum and the opposite of the \Bzb momentum in the \Dstarp rest frame. The polarisation factor $F_L^{D^*}$ is sensitive to NP configurations and thus can be used to constrain NP parameters~\cite{F_Dst_th}.

\begin{figure}[tbp]
    \centering
    \includegraphics[width=0.42\linewidth]{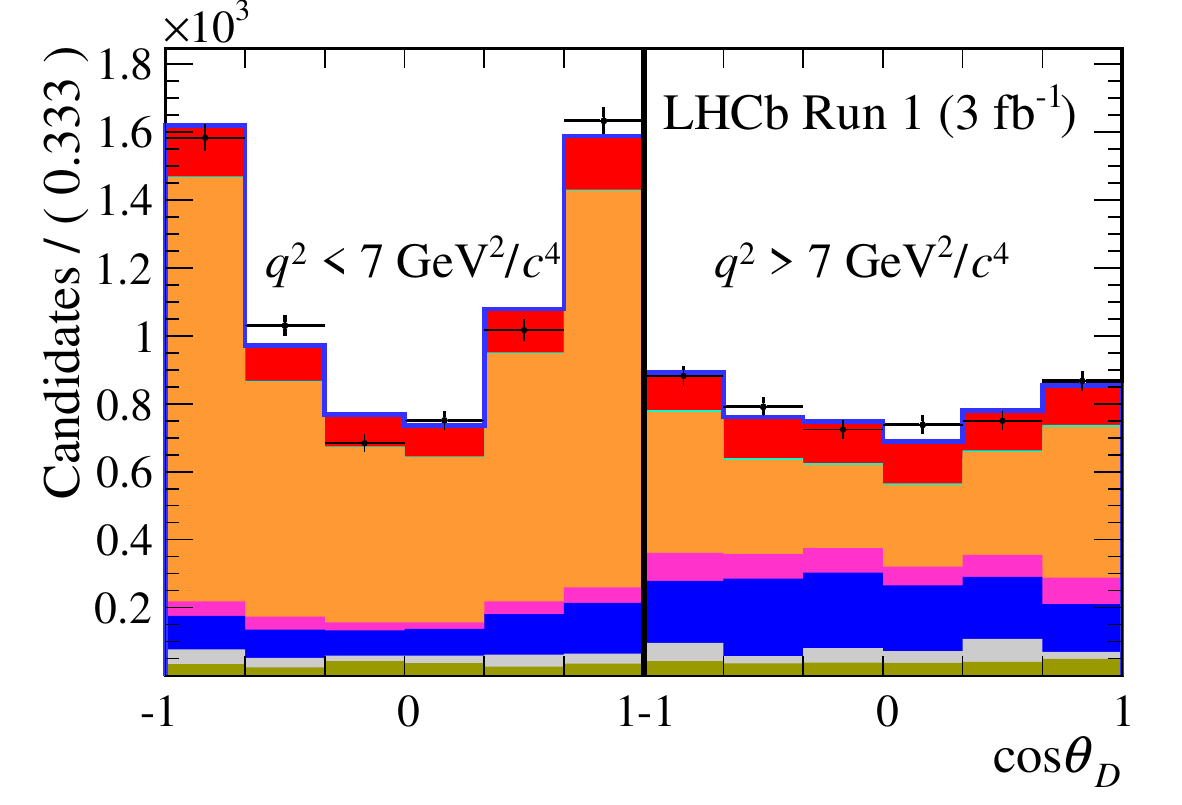}
    \includegraphics[width=0.42\linewidth]{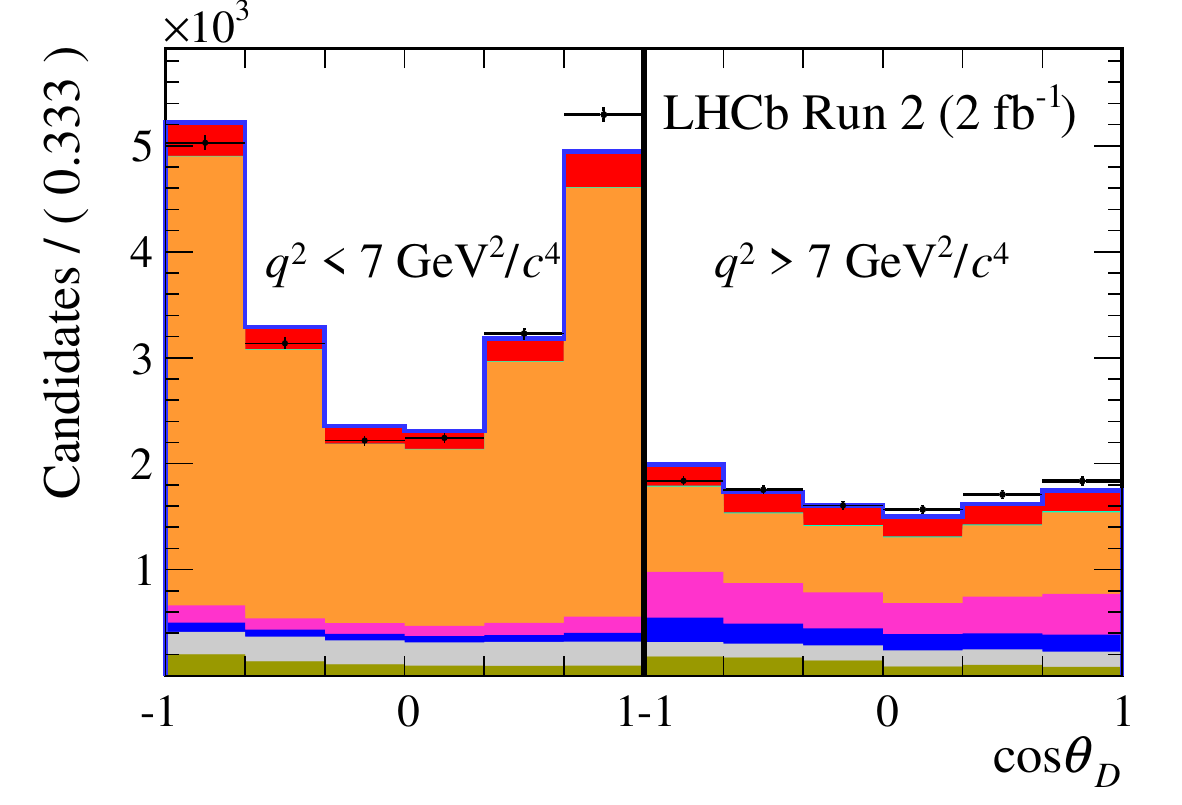}
    \raisebox{1cm}{\includegraphics[width=0.12\linewidth]{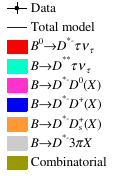}}
    \caption{The $\cos \theta_D$ distributions in two $q^2$ regions for the (left)~Run\,1 and (right)~Run\,2 $\Bz\to\Dstarm\taup\neut$ data samples with the fit results overlaid.}
    \label{fig:dstar_pol}
\end{figure}

The LHCb experiment performed a measurement of $F_L^{D^*}$ using $pp$ collision data collected in the periods 2011-2012 and 2015-2016~\cite{LHCb-PAPER-2023-020}. The $\taum$ candidates are reconstructed using the hadronic modes $\taum\to\pim\pip\pim(\piz)\neut$ and the \Dstarp are required to decay as $\Dstarp\to\Dz\pip$ followed by $\Dz\to\Km\pip$.
The coefficients $a_\theta(q^2)$ and $c_\theta(q^2)$ are extracted from a four-dimensional binned template fit to $q^2$, $\cos \theta_D$, \taum lifetime $t_\tau$ and anti-$D_s$ BDT output~\cite{BDT} that distinguishes the signal decay from the double charm background. 
The fit projections for the $\cos \theta_D$ distribution are shown in Fig.~\ref{fig:dstar_pol}.
The polarisation factor $F_L^\Dstar$ is evaluated in two $q^2$ bins and in the full region: 
\begin{equation}
\begin{aligned}
q^2<7 \mathrm{GeV}^2 / c^4:\, & 0.51 \pm 0.07 \pm 0.03, \\
q^2>7 \mathrm{GeV}^2 / c^4:\, & 0.35 \pm 0.08 \pm 0.02, \\
q^2 \text { integrated}:\, & 0.43 \pm 0.06 \pm 0.03,
\label{eq:dst_pol_result}
\end{aligned}
\end{equation}
where the first uncertainty is statistical and the second systematic. The primary systematic uncertainties stem from the limited size of simulation samples, parameterisation of the form factor and modelling of double charm background.
The results in Eq.~\ref{eq:dst_pol_result} are the most precise to date and are compatible with the Belle result~\cite{F_Dst_Belle} and the SM prediction~\cite{F_Dst_SM}.

\section{Conclusion and prospects}
In summary, this report presents two recent LHCb measurements related to $B$ semileptonic decays. The results of $R(\Dp)$ and $R(\Dstarp)$, obtained from the first LHCb analysis using the $\Dp\to\Km\pip\pip$ mode to measure $R(D^{(*)})$, align with the previous world averages and the SM expectations~\cite{HFLAV21,LHCb-PAPER-2024-007}. The updated world averages with these new results included are slightly closer to the SM predictions but the tension persists at a $3\sigma$ level. 
The first LHCb measurement of the \Dstarp longitudinal polarisation in the $\Bzb\to\Dstarp\taum\neutb$ decay offers the most precise result to date, which is compatible with the SM prediction~\cite{LHCb-PAPER-2023-020,F_Dst_SM}. 
These measurements contribute to the continuous effort of testing the SM and searching for NP in semileptonic decays of $b$-hadrons. 

There are numerous physics topics related to $B$ semileptonic decays to explore with ongoing data collections at LHCb. The $R(D)$-$R(D^{*})$ tension is anticipated to be better understood through more precise measurements using larger dataset, as uncertainties typically scale with luminosity.
Full angular analyses can be performed in some decays with high statistics, where different angular coefficients are sensitive to different NP scenarios. 
Additionally, as the accessible dataset grows, it becomes feasible to investigate decays with low statistics, such as  those involving excited charm hadrons. Measurements of these decays can provide new insights for NP searches.

\end{document}